\documentclass[11pt]{llncs}
\usepackage[letterpaper,hmargin=1in,vmargin=1.25in]{geometry}

\usepackage[applemac]{inputenc}

\usepackage{amssymb,amsmath}

\newcommand{\p}{^{\prime}}
\newcommand{\mdag}{^{\dag}}

\newcommand{\ensemble}[1]{\{ #1 \}}

\newcommand{\ket}[1]{\left|#1\right\rangle}

\newcommand{\bra}[1]{\left\langle#1\right|}

\newcommand{\braket}[2]{\left\langle#1|#2\right\rangle}
\newcommand{\densite}[1]{\left|#1\right\rangle\left\langle#1\right| 
}
\newcommand{\mtr}[1]{\mathrm{Tr}\left(#1\right)}

\newcommand{\mh}{\mathrm{H}}

\newcommand{\msf}{\mathsf}
\newcommand{\mrm}{\mathrm}

\newcommand{\mbb}{\mathbb}
\newcommand{\mcal}{\mathcal}

\newcommand{\mbI}{\mathbb{I}}
\newcommand{\mE}[1]{\mcal{E}(#1)}

\newcommand{\mpr}{{\mrm{Pr}}}

\begin{document}
\title{Entropic security in  Quantum Cryptography}
\author{Simon Pierre Desrosiers}
\institute{ Universit\'e McGill, \\ Montr\'eal, Qu\'ebec}
\maketitle
\begin{abstract}
We present two new definitions of security for quantum ciphers which are inspired by the definition of entropic security and entropic indistinguishability defined by Dodis and Smith.  We prove the equivalence of these two new definitions.   We also propose a generalization of a cipher described by Dodis and Smith and show that it can actually encrypt $n$ qubits using less than $n$ bits of key under reasonable conditions and yet be secure in an information theoretic setting.  This cipher also totally closes the gap between the key requirement of quantum ciphers and classical ciphers.
\end{abstract}
\section{Introduction}
In a seminal paper, Goldwasser and Micali \cite{GM1984}, proposed a new definition of security in classical cryptography.  In this article, they gave two new definitions of security, semantic security and indistinguishability, and proved that they are in fact equivalent.  This was a significant paradigm shift from previously used security definitions.  Both definitions rely on limitations imposed on an adversary that would intercept a cipher text, that is, the adversary is limited to be a probabilistic polynomial-time machine.  This is fundamentally a computational  security setting.  How to translate these definitions to the information setting remained unknown for years.  Russell and Wang \cite{RW2002} introduced in 2002 a satisfactory definition for the information theoretic setting.  In their paper, Russell and Wang shifted the limitations imposed to the adversary from computational limitations  to entropic limitations:  the adversary limitation is in fact a lower bound on its min-entropy on the message space, bound which we will denote by $t$.  It is to be mentioned that similar concepts for hash functions had already been developed by Canetti et al. in \cite{canetti97,CMR1998}.  Unfortunately, Russell and Wang had limited the scope of their definition by requiring that the adversary could not predict any predicate of the message based on the cipher text.  Furthermore, their proof was rather involved.  This was remedied by Dodis and Smith \cite{DS2004}  who extended the definition to all functions, gave a new definition of entropic indistinguishability similar to that of Goldwasser and Micali and proved the equivalence between the two.  They also provided three different encryption schemes and proofs that they are secure according to entropic security.  Entropic security is, in a sense, surprising since it is allowing relaxation of the traditional informational theoretic security definition that goes back to Shannon.  Indeed, if one requires that the mutual information between the cipher text and the message be smaller than some $\epsilon$, then one can only shorten the key length by $\epsilon$ bits.  Entropic security lets one reduce the key size to $n-t+2\log{(1/\epsilon)}+\mcal{O}(1)$ bits \cite{DS2004}, which constitutes a nontrivial improvement.  If the reader whishes to gain some intuition on why this is possible, we advise him to read the introduction of \cite{DS2004}.

In parallel to this development, quantum security began with a Shannon-like definition of security which requires the cipher text $\mE{\rho}$ to be equal to $\rho_{0}$, some fixed state, for all messages $\rho$ in the message space.  This was initially proposed by Ambainis, Mosca, Tapp and de Wolf in 2000 \cite{amtw00}.  This definition was later relaxed by Hayden, Leung, Shor, Winter \cite{HLSW2004} by requiring that the distance between the cipher text and the perfectly mixed state be smaller than some security parameter $\epsilon$.  They also made the critical assumption that the eavesdropper  was not entangled with the sender, an assumption which was not necessary in \cite{amtw00} --- we also impose this condition in order for our scheme to work.  Unfortunately, their proof was not constructive, but they proved that there exists an encryption scheme such that the key length is $n+\log(n)+2\log(1/\epsilon)$.  Ambainis and Smith \cite{AS2004} then gave explicit polynomial-time constructions that can reduce the key space further in certain conditions.  Their first construction uses $n+2\log(n)+2\log(1/\epsilon)+\mcal{O}(1)$ bits of key to encrypt $n$ qubits.  Their second construction, which is not length preserving, even goes down to $n+2\log(1/\epsilon)$ bits of key per $n$ qubits.  

In this paper we shall reduce the key size even further.  We shall prove that a generalization  of a scheme proposed by Dodis and Smith can  actually use less key, that is $n-t+2\log(1/\epsilon)$ bits of key for $n$ qubits, where $t$ represents the min-entropy of the adversary on the message space.  This means that if $t$ is greater than 
$2\log(1/\epsilon)$, which is not unreasonable, then the scheme actually requires less than $n$ bits of key for $n$ qubits.  We achieve this by generalizing the entropic security and indistinguishability definitions contained in \cite{DS2004} to a quantum setting.  We also prove their equivalence.  These definitions look quite simple and straightforward, but in fact are quite unsettling to anyone close to the quantum computing community.  The paper is divided as  follows:  in Section \ref{defini} we present our definitions of security and discuss the details of their interpretation.  In Section \ref{para} we prove that they are equivalent.  In Section \ref{cipher} we present an encryption scheme and prove that it uses $n-t+2\log(1/\epsilon)$ bits of key for $n$ qubits. Further contributions of this paper are: 1) a much simpler proof of the equivalence for all functions between our two definitions; 2) the first information-theoretic quantum encryption scheme which does not require more key than its classical counterpart.  We assume that the reader is well-familiar with quantum information theory, that is linear algebra, POVMs, super-operators,  distance measures and operator decompositions.  For an introduction see \cite{NC2000}.

\section{Definitions}\label{defini}

We are interested in the following scenario.  The sender chooses a message from a known message space and encrypts the message.  We want that whenever an adversary intercepts an encryption it can not predict any function on the message.  More formally, let an {\bf interpretation} of  $\rho=\sum_j \gamma_j\densite{j}$ be an ensemble $\{ (p_i,\sigma_i) \}$ such that $\rho = \sum_i p_i\sigma_i$, we say $\sigma_{i}$ is {\bf compatible\label{compatible}} with $\rho$.  This is the eavesdropper's view of the message space --- {\emph i.e.} the a priori knowledge of the adversary is given by the ensemble $\{ (p_i,\sigma_i) \}$, which consists of all the possible messages (by which we mean valid density operators, or physically possible messages) with non-zero probability along with their probability.  We want that whenever the sender chooses a message $\sigma_{i}$ and encrypts it using a cipher $\mcal{E}$, then no eavesdropper which intercepts $\mcal{E}(\sigma_{i})$ can guess any function of $\sigma_{i}$.  We will require  this property to hold for all $\rho$ with sufficiently high min-entropy, where $H_{\infty}(\rho)\triangleq H_{\infty}(\{\gamma_{j}\})\triangleq - \log(\max_{j}(\gamma_{j}))$.  Naturally, this question is pertinent in a context where interaction is not desirable or too costly to use.  

\begin{definition}\label{cipherdefinition}
An operator $\mcal{E}$ is an encryption scheme if there exists another operator $\mcal{D}$ such that for all states $\rho$ we have $\mcal{D}(\mcal{E}(\rho))=\rho$.
\end{definition}

\begin{definition}\label{sec-ent-faible}
An encryption scheme $\mcal{E}$ is said to be $(t,\epsilon)$-entropically secure if for all states $\rho$ such that $H_{\infty}(\rho)\geqslant t$, all associated interpretations $\{ (p_i,\sigma_i) \}$, and all adversaries $A$ there exists $A\p$ such that for all functions $f$ we have  
\begin{equation}
\left|\mrm{Pr}_{i}[A(\mcal{E}(\sigma_{i}))=f(\sigma_{i})] -\mrm{Pr}_{i}[A\p(\cdot)=f(\sigma_{i})] \right| < \epsilon. \footnote{Note that probabilities are not taken only over $i$, but also all randomness used by $A$, $A\p$ and $\mcal{E}$. }
\end{equation}
\end{definition}
A few explanations are in order.  First, in this equation, only one state is physical, that is, $\mcal{E}(\sigma_{i})$.  For this equation to be meaningful, all other states are not considered to be physical but purely mathematical.  By this we mean that the $\sigma_{i}$ are considered to be strings of bits that can be interpreted as density operators. This is reasonable since $A\p$ never gets his hands on any ciphers, exactly as in the traditional definition.  Hence, $\{ (p_i,\sigma_i) \}$ simply is the a priori knowledge of $A$ and $A\p$ on the message space from which the sender samples.  We therefore naturally consider that the output of $f(\sigma_{i})$ is simply a string of bits.\footnote{Note that instead of considering functions on strings of bits, one could consider that the function $f$ acts on the indices $i$ of $\sigma_{i}$ and get an equivalent framework.}  Furthermore we do not impose any restriction on $f$.  In particular, we do not require that $f$ be a physical process, hence $f$ is not required to be linear or to be a function on operators --- $g(\rho)=\sum_{i}g(\gamma_{j})\densite{j}$.

Hence $A$ has to predict the output of the function $f$ on the string of bits that represents the state $\sigma_{i}$, which is unknown to $A$, by only analyzing $\mcal{E}(\sigma_{i})$ which is a physical state --- no restriction are put on $A$,  we only require it to be a physical process, {\emph i.e} a $\mrm{POVM}$.  The adversary $A\p$ does not get this chance, he must predict the same function $f$ on the same bit string but having access to nothing but the message interpretation $\{(p_{i},\sigma_{i})\}$.  The obvious best strategy for $A\p$ is to bet on the most probable output for $f$, since all other outputs have less chance of occurring.  So by definition $\mpr_{i}[A\p(\cdot)\mbox{ is right}]=\mrm{Max}_{f}\triangleq\max_{z}\mrm{Pr}_{i}[f(\sigma_{i})=z]$ where $Z=\{z\}$ is the set of possible outputs for the interpretation $\{(p_{i},\sigma_{i})\}$ --- note that we assume that $A$ and $A\p$ know the correct interpretation which is considered to be the message space.  Quantum entropic security states that if $A$ can predict the function $f$ with a given probability, then this probability can be matched by $A\p$ up to $\epsilon$, equivalently $\mrm{Pr}_{i}[A(\mcal{E}(\sigma_{i}))=f(\sigma_{i})] \leqslant \mrm{Max}_{f}+\epsilon$.

As in \cite{GM1984} and \cite{DS2004}, we can introduce a notion of indistinguishability and then show that indistinguishability and entropic security are equivalent.

\begin{definition}\label{indis}
An encryption scheme $\mcal{E}$ is said to be $(t,\epsilon)$-indistinguishable if for all operators $\rho$ such that $H_{\infty}(\rho)\geqslant t$ we have
\begin{equation}
 \left\|\mcal{E}(\rho)-\frac{\mbb{I}}{d} \right\|_{\mrm{tr}}<\epsilon,
\end{equation}
where $d$ is equal to the dimension of the message space.
\end{definition}

We use the following definition for the trace distance: $\left\| \rho-\sigma\right\|_{\mrm{tr}}=\frac{1}{2}\mrm{tr}\left|\rho-\sigma\right|$ and $|A|\triangleq \sqrt{A\mdag A}$.  The following theorem, Theorem 9.1 in \cite{NC2000}, will be useful.  
\begin{theorem}\label{unelement}
Let $\{E_{m}\}$ be a POVM with $p_{m}=\mtr{E_{m}\rho}$ and $q_{m}=\mtr{E_{m}\sigma}$ as the probabilities of obtaining a measurment outcome labeled by $m$.  Then
\begin{equation}
D(\rho,\sigma)=\max_{\{E_{m}\}}D(p_{m},q_{m})=\max_{\{E_{m}\}}\left\{\frac{1}{2}\sum_{m}\left|\mtr{E_{m}(\rho-\sigma)}\right|\right\},
\end{equation}
where the maximization is over all POVMs $\{E_{m}\}$ and $D(\rho,\sigma)\triangleq \frac{1}{2}\left\| \rho-\sigma\right\|_{\mrm{tr}}$.
\end{theorem}
It is now time to introduce one assumption we are making all along this article.  We assume that $\mE{{\mbb{I}/d}}=\mbb{I}/d$, which is true for all reasonable schemes we know: our cipher in section \ref{cipher} will have this property. This is very powerful since we can, by the spectral decomposition theorem, decompose $\mbI/d$ in the basis of our choice, in particular the basis of $\mE{\rho}$.  This is equivalent to saying that $\mbI/d$ and $\mE{\rho}$ commute (which is trivially true for $\mbb{I}/d$), hence that the trace distance between the two operators is equal to the statistical distance of their eigenvalues: $\left\|\mE{\rho}-\mbb{I}/d\right\|_{\mrm{tr}}=D(\lambda_i,\frac{1}{d})$. 
This basically means that $(t,\epsilon)$-indistinguishability implies that there is no $\mrm{POVM}$ that can distinguish between $\mE{\rho}$ and $\mbI/d$.\footnote{See \cite{NC2000} Chapter 9 for a proof these statements.} 

It is also easy to see, using the triangle inequality, that Definition \ref{indis} implies:
\begin{definition}\label{indis2}
An encryption scheme $\mcal{E}$ is said to be weakly $(t,\epsilon)$-indistinguishable if for all operators $\rho$ and $\rho\p$ such that $H_{\infty}(\rho)\geqslant t$ and $H_{\infty}(\rho\p)\geqslant t$   we have
\begin{equation}
 \left\|\mcal{E}(\rho)-\mcal{E}(\rho\p) \right\|_{\mrm{tr}}< 2\epsilon.
\end{equation}
\end{definition} 
Obviously weak $(t,\epsilon)$-indistinguishability implies $(t,2\epsilon)$-indistinguishability. We must introduce a fourth notion which is a strong version of Definition \ref{sec-ent-faible}.

\begin{definition}\label{strong}
An encryption scheme $\mcal{E}$ is said to be strongly $(t,\epsilon)$-entropically secure if for all states $\rho$ such that $H_{\infty}(\rho)\geqslant t$, all interpretations $\{ (p_i,\sigma_i) \}$ and all adversaries $A$ we have for all function $f$
\begin{equation}\label{85}
\left|\mrm{Pr}_{i}[A(\mcal{E}(\sigma_{i}))=f(\sigma_{i})] -\mrm{Pr}_{i}[A(\mE{\rho})=f(\sigma_{i})] \right| < \epsilon .
\end{equation}
\end{definition}
The only difference with Definition \ref{sec-ent-faible} is that we have restricted the notion of $A\p$: this adversary is now the same as $A$ but it receives an encryption of $\rho$.  Basically, Equation \eqref{85} means that whatever $A$ can compute from $\mE{\sigma_{i}}$, with probability up to $\epsilon$ he could have computed it using only an oracle serving an encryption of $\rho$ which is totally independent of $\sigma_{i}$.  This strategy is clearly worse that the optimal one, since 
\begin{equation}
\mrm{Pr}_{i}[A(\mcal{E}(\sigma_{i}))=f(\sigma_{i})] \leqslant \mrm{Pr}_{i}[A(\mE{\rho})=f(\sigma_{i})] +\epsilon\leqslant \mrm{Max}_{f}+\epsilon\mbox{,}
\end{equation}
because no strategy can do better than $\mrm{Max}_{f}$ without seeing the cipher text.
  Therefore, it means that $\mE{\cdot}$ must be a better scheme to achieve this definition, since $A$ can not do much better than this worse strategy can do ($\epsilon$ better at best).  Furthermore, not only does there exist an adversary $A\p$ but this adversary can be easily constructed using $A$ as a black box.  Hence strong-entropic security implies entropic security.



As is traditionally the case in semantic security, Definition \ref{sec-ent-faible} carries the meaning of what is considered a secure encryption scheme.  Definition \ref{indis} will allow us to prove that a given scheme is secure and Definition \ref{strong} will let us show easily that these two definitions are equivalent for all functions and not just for predicates.


\section{Equivalence between the two paradigms }\label{para}

\begin{lemma}\label{lem100}
Strong $(t,\epsilon)$-entropic security implies $(t-1,2\epsilon)$-indistinguishability as long as $t\leq n-1$.
\end{lemma}
 {\bf Proof :} \\
We are translating, for this lemma, the proof from Dodis and Smith to the quantum setting.  The last part, for non-orthonormal states, is new to this work. It is well known that a classical  $t$-source\footnote{A $t$-source is a random variable with min-entropy no less than $t$.} can be decomposed into a convex combination of flat sources over $2^{t} $ points\footnote{A $t$-flat source, is a uniform distribution over $2^{t}$ points.}.  Moreover the two are linked in an easy way: if $X$ is a classical $t$-source, and $Y$ is an equiprobable distribution on the first $2^{t}$ points (the order is arbitrary), then there exists $\{ P_{i}\}$ such that $X=\sum_{i}p_{i}P_{i}Y$, where $\sum_{i}p_{i}=1$ and the $P_{i}$'s are permutation matrices.

It is less known, yet also true, that we can say the same thing about density operators.  Let $\rho$ be a state such that $H_{\infty}(\rho)\geqslant t$ and let $\sigma$ be a perfectly mixed state with $H_{\infty}(\sigma)=H(\sigma)=t$ ({\emph i.e} the $\mrm{support}(\sigma)\mbox{ has size }2^{t}$).  Then we can decompose $\rho$ this way
\begin{equation}
\rho=\sum_{i}p_{i}U_{i}\sigma U_{i}\mdag,
\end{equation}
where $\sum_{i}p_{i}=1$ and the $U_{i}$'s are unitary operators.  It must also be said that if $\rho$ and $\sigma$ commute, then the $U_{i}$'s are just permutation matrices.\footnote{For a proof of all these statements, read the section on majorization theory in \cite{NC2000}: Section 12.5.1.}

These observations will allow us to prove the lemma for flat sources of entropy $t-1$ only.  Indeed, $\mcal{E}$ can not decrease entropy, so $H_{\infty}(\mE{\rho})\geqslant t$ and $\mbI/d$ is of course a $t$ source.  So we can write $\rho=\sum_{i}p_{i}X_{i}$ where $X_{i}=U_{i}\sigma U_{i}\mdag$, the $U_{i}$'s are permutation matrices and $\sigma$ is a flat $(t-1)$-source which we choose in the eigenbasis of $\rho$.  Similarly, we can write $\mbI/d=\sum_{j}q_{j}Y_{j}$, where $Y_{j}=V_{j}\sigma V_{j}\mdag$ (the reader should keep in mind that we can diagonalize $\mbI$ in the basis of our choice, so we choose the eigen-basis of $\rho$, hence $\rho$, $\mbI/d$ and $\sigma$ all commute with one another).  We know that $\mE{\mbI/d}=\mbI/d$. 

So $\left\| \mcal{E}(\rho)-\mcal{E}(\mbb{I}/d) \right\| = \left\|  \mcal{E}(\sum_ip_i X_i)-\mcal{E}(\sum_j q_jY_j) \right\|$.\footnote{We have dropped momentarily the $\mrm{tr}$ indices to the trace distance for compactness.} Since $\sum_i p_i = \sum_j q_j = 1$, we can write this: 
$\left\| \mcal{E}((\sum_j q_j)\sum_ip_i X_i)- \mcal{E}((\sum_i p_i)\sum_j q_jY_j)  \right\|$ which simplifies to  $\left\|  \mcal{E}(\sum_{i,j}p_iq_j X_i)-\mcal{E}(\sum_{i,j}p_i q_jY_j )\right\|\mbox{.}$
Since $\mcal{E}$ is a linear operator we can rewrite everything this way:  $\left\|  \sum_{i,j}p_iq_j\mcal{E}( X_i) - \sum_{i,j}p_i q_j\mcal{E}(Y_j )\right\|$ which we can simplify  to $\left\|  \sum_{i,j}p_iq_j(\mcal{E}( X_i) -\mcal{E}(Y_j ))\right\|.$  Using the triangle inequality, we can conclude:
\begin{equation}\label{108}
\left\| \mcal{E}(\rho)-\mcal{E}(\mbb{I}/d) \right\| \leqslant \sum_{i,j}p_iq_j\left\| \mcal{E}(X_i)-\mcal{E}(Y_j) \right\|\mbox{.}
\end{equation}
This equation tells us that if all terms $\left\| \mcal{E}(X_i)-\mcal{E}(Y_j) \right\|$ are less than $2\epsilon$, then Equation \eqref{108} is bounded by $2\epsilon$ and in particular the cipher must be $(t-1,2\epsilon)$-indistinguishable. 

So let $W_{0}\in \{X_{i}\}$, where, $\rho=\sum_{i}p_{i}X_{i}$ and let $W_{1}\in \{Y_{j}\}$.  Assume for now  they have orthonormal support.  Consider the operator $Z=1/2W_{0}+1/2W_{1}$: an equal mixture of the 2 states.  By construction, $H_{\infty}(Z)=t$.   Define the predicate $g$ such that for any state $\tau_{0}$ compatible (see page \ref{compatible}) with $W_{0}$, $g(\tau_{0})=0$, and for any state $\tau_{1}$ compatible with $W_{1}$, $g(\tau_{1})=1$.  It is not necessary to define the value of $g$ for any other state.  Any adversary, $A$, that can predict $g$ given $\mE{\tau_{b}}$, for $b\in_{R}\{0,1\}$, is therefore a distinguisher between $\mE{W_{0}}$ and $\mE{W_{1}}$.

It is common knowledge (see \cite{NC2000}) that, at  best, such an adversary can do this with probability:
\begin{equation}\label{114}
\mpr[A(\mE{\tau_{b}})=g(\tau_{b})=b]=\frac{1}{2}+\frac{1}{2}\left\|\mE{W_{0}}-\mE{W_{1}} \right\|_{\mrm{tr}}\mbox{.}
\end{equation} 
We can now invoke the entropic security definition.  So we can also write:
\begin{equation}\label{115}
\mpr[A(\mE{\tau_{b}})=g(\tau_{b})=b]\leqslant\mpr[A\p(\cdot)=g(\tau_{b})=b]+\epsilon=\frac{1}{2}+\epsilon\mbox{.}
\end{equation} 
By construction no adversary $A\p$ can guess the correct answer with probability better than one half.  Using Equations \eqref{114} and \eqref{115}, we can conclude:
\begin{equation}
\left\| \mcal{E}(W_0)-\mcal{E}(W_1) \right\|_{\mrm{tr}}\leqslant 2\epsilon\mbox{.}
\end{equation}
We are almost done.  Let us now suppose that $W_{0}$ and $W_{1}$ are not orthogonal but not equal.  Let $V$ be the space spanned by their intersection.  This space, $V$ is well defined and $V$ is not equal to $W_{0}$ nor $W_{1}$.  Because $\rho$ and $\mbI/d$ commute and because we already concluded that $W_{0}$ and $W_{1}$ commute, we can treat these objects as classical distributions on classical points, and treat their eigen-basis as points in sets (We abuse notation in that spirit here).  Hence we can create a new state $W_{0}\p$ such that $W_{0}\cap W_{0}\p = W_{0}\setminus V$ and $W_{0}\p\cap W_{1}=\emptyset$.  We can do this since $t\leqslant n-1$, so we have plenty of space to choose new points.  Of course  we choose $W_{0}\p$ such that it has min-entropy equal to $t-1$ and such that it is a $t-1$ flat source.  Obviously, by construction we have $\left\| W_{0} - W_{1}\right\| \leqslant \left\| W_{0}\p - W_{1}\right\|$, hence, using our argumentation for orthogonal states, and the fact that $W_{0}$, $W_{1}$ and $W_{0}\p$  commute, we can conclude that
\begin{equation}
\left\| \mcal{E}(W_0) - \mcal{E}(W_1) \right\|_{\mrm{tr}}\leqslant \left\| \mcal{E}(W_0\p) - \mcal{E}(W_1) \right\|_{\mrm{tr}} \leqslant 2\epsilon\mbox{.}
\end{equation}
{\sf QED}.
\begin{lemma}\label{lem200}
Let $\rho$ be a state such that $H_{\infty}(\rho)\geqslant t$ and let $\{ (p_i,\sigma_i) \}$ be an interpretation of $\rho$.  Then for all $i$ we have that $p_{i}\cdot \lambda_{max_i} \leqslant 2^{-t}$, where $\lambda_{max_i}$ is the biggest eigenvalue of $\sigma_{i}$.
\end{lemma}
{\bf Proof :} \\
Suppose, on the contrary, that $p_{i}\cdot\lambda_{max_i}>2^{-t}$.  Since $\lambda_{max_i}$ is an eigenvalue of $\sigma_{i}$, there exists a vector $\ket{v}$ such that $\bra{v}\sigma_{i}\ket{v}=\lambda_{max_i}$.  These two observations together let us write $\bra{v}\rho\ket{v}\geqslant p_{i}\bra{v}\sigma_{i}\ket{v}>2^{-t}$.  We also know that $\rho=\sum_{k}\gamma_{k}\densite{k}$, so  $\bra{v}\rho\ket{v}=
\sum_{k}\gamma_{k}\braket{v}{k}\braket{k}{v}\leqslant \sum_{k}2^{-t}\braket{v}{k}\braket{k}{v}=2^{-t}$.  Hence we conclude that $2^{-t}< \bra{v}\rho\ket{v} \leqslant 2^{-t}$, which is an obvious contradiction.\\
{\sf QED}.
\begin{lemma}\label{allfunction}
Let $\rho$ be a state, $\{ (p_i,\sigma_i) \}$ be an interpretation, $\mcal{E}$ be a cipher, $f$ be a function and $A$ be an Adversary such that 
$$\left|\mrm{Pr}_{i}[A(\mcal{E}(\sigma_{i}))=f(\sigma_{i})] -\mrm{Pr}_{i}[A(\mE{\rho})=f(\sigma_{i})] \right|\geqslant \epsilon,$$
then there exist an adversary $B$ and a predicate $h$ such that
$$\left|\mrm{Pr}_{i}[B(\mcal{E}(\sigma_{i}))=h(\sigma_{i})] -\mrm{Pr}_{i}[B(\mE{\rho})=h(\sigma_{i})] \right|\geqslant \frac{\epsilon}{2}.$$
\end{lemma}
{\bf Proof :} \\
Let our predicate be a Goldreich-Levin predicate, that is $h_{r}(x)=r\odot f(x)$, where $\odot$ denotes the scalar product of the binary vectors represented by the strings $f(x)$ and $r$.  Let $p=\mpr_{i}[A(\mcal{E}(\sigma_{i}))=f(\sigma_{i})]$ and $q=\mpr_{i}[A(\mcal{E}(\rho))=f(\sigma_{i})]$.  Then we know that $|p-q|\geq \epsilon$.  Let us compute
 \begin{equation}\label{allfunction1}
E= \left|\mbb{E}_{r}\left[\mrm{Pr}_{i}[r\odot A(\mcal{E}(\sigma_{i}))=h_{r}(\sigma_{i})] -\mrm{Pr}_{i}[r\odot A(\mE{\rho})=h_{r}(\sigma_{i})] \right]\right|,
 \end{equation}
where the expectation is taken over all $r$ of adequate size.
We need two observations.  First, when $A$ predicts correctly, then $\mrm{Pr}_{i}[r\odot A(\mcal{E}(\sigma_{i}))=h_{r}(\sigma_{i})]=1$.  Second, when $A$ does not predict correctly, the probability that $r\odot A(\mcal{E}(\sigma_{i}))=h_{r}(\sigma_{i})$ is exactly one half.
Hence Equation \eqref{allfunction1} reduces to
\begin{equation}
E=\left| 1\cdot p+\frac{1}{2}\cdot(1-p)-\left(1\cdot q+\frac{1}{2}\cdot (1-q)\right)\right|= \left| \frac{p-q}{2}\right| \geqslant \frac{\epsilon}{2}.
\end{equation}
There exists at least one value $r$ such that the following is true:
 \begin{equation*}\label{allfunction4}
\left|\mrm{Pr}_{i}[r\odot A(\mcal{E}(\sigma_{i}))=h_{r}(\sigma_{i})] -\mrm{Pr}_{i}[r\odot A(\mE{\rho})=h_{r}(\sigma_{i})] \right| \geqslant \frac{\epsilon}{2}.
 \end{equation*}
The lemma is proven if we define adversary $B(\cdot)$ as $r\odot A(\cdot)$ for this appropriate $r$.\\
 {\sf QED}.
\begin{lemma}\label{lemma400}
$(t-1,\epsilon/8)$-indistinguishability implies strong $(t,\epsilon)$-entropic security for all functions as long as $t\leqslant n-1$.
\end{lemma}
{\bf Proof :} \\
The proof technique used in this lemma is new to this work, as far as we know.  \\
Suppose that there exists an adversary $B$, a state $\rho$, where $H_{\infty}(\rho)\geqslant t$, an interpretation $\{ (p_j,\sigma_j) \}$ for $\rho$ and a function $f$ such that 
\begin{equation}\label{semantiquediff}
\left| \mpr_{i} \left[B(\mcal{E}(\sigma_{i}))=f(\sigma_{i})  \right] - \mrm{Pr}_{i}[B(\mE{\rho})=f(\sigma_{i})] \right| \geqslant \epsilon\mbox{.}
\end{equation}
We want to show that this adversary implies that the encryption scheme $\mcal{E}$ is not $(t-1,\epsilon/8)$-indistinguishable.
Then, by the previous lemma, we know that there exists another adversary  $A$ and a predicate $h$ such that strong $(t,\epsilon/2)$-entropic security is violated. 
Let us define two sets $E_{0}$ and $E_{1}$ this way: \\[-5ex]
\begin{itemize}
\item $E_{0}=\{ i | h(\sigma_{i})=0 \}$
\item $E_{1}=\{ i | h(\sigma_{i})=1 \}$.
\end{itemize}
Let $r_{0}=\sum_{i\in E_{0}}p_{i}$ and $r_{1}=\sum_{i\in E_{1}}p_{i}$.  Let $\tau_{0}=\left(\sum_{i\in E_ {0}}p_{i}\sigma_{i}\right)/r_{0}$ and $\tau_{1}=\left(\sum_{i\in E_ {1}}p_{i}\sigma_{i}\right)/r_{1}$.  Obviously, $\rho$ is equal to $r_{0}\tau_{0}+r_{1}\tau_{1}$ and both $\tau_{0}$ and $\tau_{1}$ are valid density operators.  
So if we restate the entropic security violation in terms of the $\tau_{i}$, we get
\begin{equation}\label{entsecviolation}
\left| \mpr_{i}\left[A(\mcal{E}(\tau_{i}))=h(\tau_{i})\right] - \mpr_{i}\left[A(\mcal{E}(\rho))=h(\tau_{i})\right] \right|\geqslant \frac{\epsilon}{2},
\end{equation}
where $h(\tau_{i})=i$.
The adversary $A$ is a POVM with two elements --- $A_{0}$ and $A_{1}$---, so we can rewrite equation \eqref{entsecviolation} this way:
\begin{equation}
\left| \sum_{i=0,1}p_{i}\Big(\mtr{A_{h(\tau_{i})}\mcal{E}(\tau_{i})}-\mtr{A_{h(\tau_{i})}\mcal{E}(\rho)} \Big) \right| \geqslant\frac{\epsilon}{2}
\end{equation}
where $\mtr{A_{k}\gamma}$ is the probability that $A$ on $\gamma$ outputs $k$, in our case, there are only two possible outputs: zero and one.  From the last equation, since there are only two terms in the sum, we can conclude that there exists $i$ such that 
\begin{equation}\label{contradictionassumption}
\left| p_{i}\Big(\mtr{A_{h(\tau_{i})}\mcal{E}(\tau_{i})}-\mtr{A_{h(\tau_{i})}\mcal{E}(\rho)} \Big) \right| \geqslant\frac{\epsilon}{4}.
\end{equation}
Let us assume without loss of generality that $i$ is in fact zero and let us construct the two following states (choosing $i$ to be one, would lead to a similar argument):
\begin{itemize}
\item $\tau_{0}\p \triangleq r_{0}\tau_{0 }+ r_{1}\frac{\mbI}{d} $
\item $\rho\p \triangleq r_{0}\rho + r_{1}\frac{\mbI}{d} $
\end{itemize}
Obviously, $\rho\p$ is a $t$-source since it is a convex combination of two $t$-sources.  On the other hand,  the largest eigen-value of $\tau_{0}\p$ cannot be larger than $2^{-t}+r_{1}*\frac{1}{d}$ (we have used Lemma \ref{lem200}, and the fact that we can decompose  $\mbI/d$ in the same basis as the eigen basis of $\tau_{0}$).  Since $r_{1}1/d\leqslant 2^{-t}$, we conclude that the largest eigen-value of $\tau_{0}\p$ is not larger than $2^{-(t-1)}$.  Hence, $H_{\infty}(\tau_{0}\p)\geqslant t-1 $.

Let us now compute the following expression:
\begin{equation}
\left|\mtr{A_{h(\tau_{0})}\mcal{E}(\tau_{0}\p)}-\mtr{A_{h(\tau_{0})}\mcal{E}(\rho\p)} \right|= \left|\mtr{A_{h(\tau_{0})}\big(\mcal{E}(\tau_{0}\p)-\mcal{E}(\rho\p) }\right| ,
\end{equation}
which will give us a lower bound on the trace distance of $\mcal{E}(\tau_{0}\p)$ and $\mcal{E}(\rho\p)$ as Theorem \ref{unelement} tells us, since $A_{h(\tau_{0})}$ is a fixed POVM element.
\begin{equation*}
\begin{split}
\Big| &\mtr{A_{h(\tau_{0})}\mcal{E}(\tau_{0}\p)} - \mathrm{Tr}(A_{h(\tau_{0})} \mcal{E}(\rho\p)) \Big| \\
 &= \left|\mtr{A_{h(\tau_{0})}\mcal{E}\left(r_{0}\tau_{0}+r_{1}\frac{\mbI}{d}\right)}-\mtr{A_{h(\tau_{0})}\mcal{E}\left(r_{0}\rho+ r_{1}\frac{\mbI}{d}\right)} \right|\\
 &= \left|\mtr{A_{h(\tau_{0})}\mcal{E}\left(r_{0}\tau_{0}\right)}+\mtr{A_{h(\tau_{0})}\mcal{E}\left(r_{1}\frac{\mbI}{d}\right)}-\mtr{A_{h(\tau_{0})}\mcal{E}\left(r_{0}\rho \right)}- \mtr{A_{h(\tau_{0})}\mcal{E}\left(r_{1}\frac{\mbI}{d}\right)} \right|\\
 &= \left|r_{0}\Big(\mtr{A_{h(\tau_{0})}\mcal{E}\left(\tau_{0}\right)}-\mtr{A_{h(\tau_{0})}\mcal{E}\left(\rho \right)}\Big) \right|\\
 & > \frac{\epsilon}{4},
\end{split}
\end{equation*}
where the last step comes from equation \eqref{contradictionassumption}.  Hence $\tau_{0}\p$ an $\rho\p$ constitute a violation of $(t-1,\epsilon/8)$-weak-indistinguishability, which in turns implies a violation of the $(t-1,\epsilon/8)-indistinguishability$.

{\sf QED}.

We can summarize the two last lemmas in this theorem.
\begin{theorem}
Entropic security and entropic indistinguishability are equivalent up to small variation in parameters.
\end{theorem}
\section{A quantum entropic encryption scheme} \label{cipher}
We now present a generalization of the scheme proposed in Section 3.2 of \cite{DS2004}.  The proof technique used here is new to this work and achieves slightly better results than  \cite{DS2004}.
\begin{definition}\label{def69}
Let  $\mcal{H}_{n}=\ensemble{h_{i}}_{i\in I}$ be a family of permutations over $n$ bit strings.  
Consider the event $\msf{A}=h_{i}(x)\oplus h_{i}(y)$.  We say the family $\mcal{H}_{n}$ is strongly-XOR-universal if for all $x$, $y$ and all $a\not= 0$ we have
$$\mpr_{i\leftarrow I}[\msf{A}=a]\leqslant \frac{1}{2^n}. $$
\end{definition}
The family proposed in \cite{DS2004} naturally possesses this property.  Notice that the probability of seeing $\msf{A}=a=0$ can be much larger than $1/2^{n}$: in fact it is equal to the collision probability of the input.
\begin{proposition}\label{protocol2}
Let $\mcal{H}_{2n}$ be a strongly-XOR-family of permutations.  Consider the super-operator $\mcal{E}(
\rho) = \langle i, X^aZ^b\rho Z^bX^a\rangle$ \footnote{$X^{a}Z^{b}=X^{a_{1}}Z^{b_{1}}\otimes \dots \otimes X^{a_{n}}Z^{b_{n}}$ if $a=a_{1}\dots a_{n}$ and $b=b_{1}\dots b_{n}$.}, where $i$ is chosen at random uniformly over $2n$ bit strings and $a\|b=h_{i}(k)$, where $k$ is the secret key ({$a\|b$ denotes the concatenation of the strings $a$ and $b$}).  Then $\mcal{E}$ is a quantum cipher.
\end{proposition}
\begin{theorem}
The cipher of proposition \ref{protocol2} is $(t,\epsilon)$-indistinguishable for all state $\rho$ such that $H_{\infty}(\rho)\geq t$ as long as  $\mh_{\infty}(K)+\mh_{\infty}(\rho)\geqslant n + 2\log(1/\epsilon) $.
\end{theorem}
{\bf Proof :} \\
We will use the following trick:  if $\rho$ has rank $d$ and $\mtr{\mE{\rho}^{2}}\leqslant 1/d(1+\epsilon^{2})$, then $\| \mE{\rho} - \mbI/d \|_{\mrm{tr}}\leqslant \epsilon$, which implies the desired $(t,\epsilon)$-indistinguishability.\footnote{See \cite{AS2004} and \cite{DS2004}.}
The adversary's view can be written this way: $\rho\p=\mcal{E}(\rho)=\mbb{E}_{a,b,i}[i\otimes X^aZ^b\rho Z^bX^a ]$.  We are interested in the following quantity $\mtr{\mcal{E}(\rho)^{2}}$.  So
\begin{eqnarray}
\mtr{\mcal{E}(\rho)^{2}} & = & \frac{1}{| I |}\mtr{\mbb{E}_{k,k\p,i}[ X^{a}Z^{b}\rho Z^{b}X^{a}X^{c}Z^{d}\rho Z^{d}X^{c} ]} \\
 & = &\frac{1}{| I |}\mtr{\mbb{E}_{k,k\p,i}[Z^{d}X^{c}X^{a}Z^{b}\rho Z^{b}X^{a}X^{c}Z^{d}\rho  ]} \\
 & = & \frac{1}{| I |}\mtr{\mbb{E}_{k,k\p,i}[ (-1)^{d\odot c} (-1)^{d\odot a} X^{c}X^{a}Z^{d}Z^{b}\rho Z^{b}X^{a}X^{c}Z^{d}\rho  ]}\\
 & = & \frac{1}{| I |}\mtr{\mbb{E}_{k,k\p,i}[ ((-1)^{d\odot c})^2 ((-1)^{d\odot a})^2 X^{c}X^{a}Z^{d}Z^{b}\rho Z^{b}Z^{d}X^{a}X^{c}\rho  ]}\\
 & = & \frac{1}{| I |}\mtr{\mbb{E}_{ef,i}[  X^{e}Z^{f}\rho Z^{f}X^{e}\rho  ]}\label{eq35} 
\end{eqnarray}
where $a\|b=h_{i}(k)$ and $c\|d=h_{i}(k\p)$ and where $k$ and $k\p$ are independent instances of the key.  Also $e\| f= (a\oplus c )\| (b \oplus d) = (a \|b)\oplus (c\| d)$.  By Definition \ref{def69}, we know that the probability of seeing any string $e\| f$, different from zero, is bounded above by $1/2^{2n}$.  Let us divide Equation \eqref{eq35} into two terms, one for $e\|f = 0$ and the other for all the $e\|f \not= 0$.  Let us introduce the following notations: $\rho_{ef}$ instead of $X^{e}Z^{f}\rho Z^{f}X^{e}$ and $p_{ef}$ for the probability that $e\|f $ is observed.  Thus, we can rewrite everything like this :
 \begin{equation}\label{eq45}
  \mtr{\mcal{E}(\rho)^{2}}  = \frac{1}{| I |}\mtr{\frac{\rho^{2}}{|K|} + \sum_{\substack{e,f\\ \mbox{\scriptsize where } e\|f\not=0} } p_{ef}\rho_{ef}\rho    }. 
  \end{equation}
  Observe two things: for all $e\|f\not = 0$ we know that $p_{ef}\leqslant 1/2^{2n}$ and $\sum_{ef}\frac{1}{2^{2n}}\rho_{ef}=\mbI/2^n$, the perfectly mixed state.  Quantum mechanic also tells us that $\mtr{\rho\sigma}$ is the expectation of the observed eigenvalue if one measures the observable $\rho$ on the state $\sigma$.  A specific case is $\mtr{\frac{\mbI}{2^n}\rho}=1/2^n$, since all eigenvalues of the perfectly mixed state are equal to $1/2^n$, the average can not be different from this number.
    
Let $A$ be the positive operator $\sum_{\mspace{-9mu}\substack{e,f\\  e\|f\not=0}}p_{ef}\rho_{ef}$.  From the previous observations, we can conclude that there exists a positive operator $B$ such that $A+B=\mbb{I}/2^{n}$ --- $B=\sum_{e,f}(\frac{1}{2^n}-p_{ef})\rho_{ef}$ and $p_{0\|0}=0$. Therefore $\mtr{(A+B)\rho}\leqslant \frac{1}{2^n}$, thus $\mtr{A\rho}+\mtr{B\rho}\leqslant \frac{1}{2^n}$ and finally $\mtr{A\rho}\leqslant \frac{1}{2^n}$.

So we can rewrite Equation \eqref{eq45} this way:
\begin{equation}\label{eq55}
 \mtr{\mcal{E}(\rho)^{2}}  \leqslant \frac{1}{| I |} \left( \mtr{\frac{\rho^{2}}{|K|}}+\frac{1}{2^n}     \right).  
 \end{equation}
Let us denote $\mh_{\infty}(K)$ by $t_{K}=\log |K|$ and $\mh_{\infty}(\rho)$ by $t_{\rho}$.  By hypothesis, we have $\mh_{\infty}(K)+\mh_{\infty}(\rho)\geqslant n + 2\log(1/\epsilon) $, hence $2^{n-t_{k}-t_{\rho}}\leqslant \epsilon^2$.  We can thus rewrite \eqref{eq55} this way:
\begin{equation}\label{eq65}
 \mtr{\mcal{E}(\rho)^{2}}  \leqslant \frac{1}{| I |}\frac{1}{2^n} \left( {{2^{n-t_{k}-t_{\rho}}}+1 }   \right)\leqslant \frac{1}{| I |}\frac{1}{2^n} \left( \epsilon^2+1 \right),  
 \end{equation}
 since $\mtr{\rho^2}\leqslant 1/2^{t_{\rho}}$.   This, in turn, implies that $\left\| \mcal{E}(\rho)-\frac{\mbb{I}}{d} \right\|_{\mrm{tr}}\leqslant \epsilon$ for $t_{k}=\log{|K|}\geqslant n-t+2\log(1/\epsilon)$.\\
 {\sf QED}.

\section{Discussion}
We have proposed two new definitions of security for quantum ciphers which are generalizations of entropic security and entropic indistinguishability.  We have proven the equivalence of the two definitions up to slight variations of the parameters $t$ and $\epsilon$.  We also presented the most efficient, in terms of the key size, quantum encryption scheme known yet.  It is not hard to prove that the three schemes presented by Ambainis  and Smith in \cite{AS2004} are all $(t,\epsilon)$-indistinguishable.  Furthermore, the first of these schemes, which uses $\delta$-biased spaces, was also presented as a classical entropically-scheme in \cite{DS2004}.  Surprisingly, the quantum version of this scheme does not require longer key than its classical counterpart.  So we ask (dare we conjecture ?): is entropic security a sufficient relaxation of information theoretic security so that quantum ciphers require no more key than their classical equivalent?  If so, is this the simplest such relaxation possible?
\bibliographystyle{plain}
\bibliography{entropicsecurity}

\end{document}